\documentstyle[12pt]{article}                                                  
\input epsf
\begin{document}
\title{\bf Stability Conditions and the Single Mode Approximation in FQHE}

\author{{\bf  Alejandro Cabo\thanks{Permanent Address:
Grupo de F\'{\i}sica Te\'orica, Instituto de Cibern\'etica,
 Matem\'atica y F\'{\i}sica,
Calle E, No. 309, Esq. a 15, Vedado, La Habana 4, Cuba}} \\
{\small\it International Centre for Theoretical Physics,}\\
{\small\it P.O.Box. 586,34100, Trieste, Italy}\\
\, \\
{\bf Aurora P\'erez-Martinez   }\\
{\small\it Grupo de F\'{\i}sica Te\'orica, 
Instituto de Cibern\'etica, Matem\'atica y F\'{\i}sica,}\\
{\small\it Calle E No.309 Esq. a 15, Vedado,
    La Habana 4, Cuba}}
\vspace{-3cm}
\maketitle

\begin{abstract}
\noindent
 A thermodynamic stability criterion for the spontaneous
 breaking of  translation invariance in many particle systems
 is derived. It simply  requires the positive character of the
 wavevector dependent dielectric function, generalising the
 same condition for  macroscopic dielectric constants.
 Its application to the  
 Single Mode Approximation (SMA) for the description of 
  collective modes of the $\nu=1/m$  Laughlin states is considered.
  The results indicate that the stability condition is violated by the
  SMA for all the relevant states $ m=3,5,7,9$ in
  a wavevector neighborhood of the magneto-roton  minima.
  These conclusions  are in qualitative agreement with
  similar results obtained from the solution of the  Bethe-Salpeter
 equation at $\nu=1/3$ for both composite fermions
 and phenomenologically described  electrons in the Laughlin state.

\end{abstract}
\newpage
\baselineskip=25pt

\newcommand{\xv}{\vec{x}}
\newcommand{\yv}{\vec{y}}
\newcommand{\kv}{\vec{k}}
\newcommand{\qv}{\vec{q}}
\newcommand{\mk}{\left| \vec{k} \right| }
\newcommand{\mq}{\left| \vec{q} \right| }
\newcommand{\fx}{\phi(\vec{x})}
\newcommand{\fy}{\phi(\vec{y})}
\newcommand{\fk}{\phi(\vec{k})}
\newcommand{\fq}{\phi(\vec{q})}
\newcommand{\px}{\Psi(\vec{x})}
\newcommand{\py}{\Psi(\vec{y})}
\newcommand{\phx}{\Psi^{+}(\vec{x})}
\newcommand{\phy}{\Psi^{+}(\vec{y})}
\newcommand{\pcx}{\Psi^{*}(\vec{x})}
\newcommand{\pcy}{\Psi^{*}(\vec{y})}
\newcommand{\ga}{{\delta^{2} \Gamma
   \over \delta \phi (\vec{x}) \delta \phi (\vec{y})}}
\newcommand{\gae}{{\delta^{2}
 \Gamma_{e} \over \delta \phi (\vec{x}) \delta \phi (\vec{y})}}
\newcommand{\ix}{\int^{\beta}_{0} d x_4}
\newcommand{\iy}{\int^{\beta}_{0} d y_4}
\newcommand{\ppxo}{\Psi^{+}(\vec{x},0) \Psi(\vec{x},0)}
\newcommand{\ppyo}{\Psi^{+}(\vec{y},0) \Psi(\vec{y},0)}
\newcommand{\ppx}{\Psi^{+}(\vec{x},x_4) \Psi(\vec{x},x_4)}
\newcommand{\ppy}{\Psi^{+}(\vec{y},y_4) \Psi(\vec{y},y_4)}
\newcommand{\ppcy}{\Psi^{*}(\vec{y},y_4) \Psi(\vec{y},y_4)}
\newcommand{\ppcx}{\Psi^{*}(\vec{x},x_4) \Psi(\vec{x},x_4)}
\newcommand{\dzx}{{\delta Z_e \over \delta \phi(\vec{x})}}
\newcommand{\dzy}{{\delta Z_e \over \delta \phi(\vec{y})}}
\newcommand{\deo}{\rho_o}
\newcommand{\dx}{\hat{\rho}(\vec{x})}
\newcommand{\dy}{\hat{\rho}(\vec{y})}
\newcommand{\dk}{\hat{\rho}_{\vec{k}}}
\newcommand{\dmk}{\hat{\rho}_{-\vec{k}}}
\newcommand{\ketn}{ \left| n \right> }
\newcommand{\nket}{\left< n \right| }
\newcommand{\ketm}{\left| m \right> }
\newcommand{\mket}{\left< m \right|}
\newcommand{\enen}{\epsilon_n}
\newcommand{\enem}{\epsilon_m}
\newcommand{\dif}{\mathcal{D} \Psi^{*}
      \mathcal{D} \Psi }
\newcommand{\bet}{\beta}
\newcommand{\dvd}{dx^2}
\newcommand{\dvt}{dx^3}
\newcommand{\partit}[2]{Z[#1,#2]}
\newcommand{\partet}[2]{Z_e[#1,#2]}
\newcommand{\free}[2]{\Gamma[#1,#2]}
\newcommand{\freee}[2]{\Gamma[#1,#2]}
\newcommand{\ba}{\begin{eqnarray}}
\newcommand{\ea}{\end{eqnarray}}
\newcommand{\nn}{\nonumber}

The stability properties of quantum  many body ground states
are of great interest in condensed matter theory.
Important questions such as the occurrence  of spontaneous breaking of
translation, and other, symmetries  are closely linked with issues of 
stabilty.
However, the complicated nature  of the
exact Green function equations for these problems usually  makes
 extracting information about  stability difficult.
A  particular important  example includes
the determination of the conditions for the formation
of the  Wigner crystal at low  electron gas density, both with
and without a magnetic field ~\cite{willet}, ~\cite{march1}.

In the present work, we discuss the derivation of a
requirement to  discuss the
stability of a translationally invariant many
particle system with respect to the spontaneous  breaking of this symmetry.
In short, this result is a 
generalisation of the usual positive value condition on the dielectric
constant \cite{landau} to the case of a general wavevector dependent
dielectric function in a translationally invariant many body state.

We expect this constraint may  also be helpful in rigorously
predicting  the breaking of  translation
invariance  which could be  necessary to explain 
short range order in liquids that have fully
invariant hamiltonians \cite{march2}.

We apply  the criterion  to  check 
the  so called Single Mode Approximation (SMA) calculations
of the collective mode excitations
 of the Laughlin ground states for $1/m$
values of the filling factor ~\cite{gmp}.

 The results indicate that the stability   rule
  is not satisfied  in a neighborhood of
  the magneto-roton  minima for the 
  states corresponding to $m=3,5,7,9$.
 The instability  appears  to be stronger
 at  lower filling factors as  is natural to  expect.

 These conclusions are in qualitative
 agreement with our previous results ~\cite{w1},~\cite{w2}.
 In ~\cite{w1},  the collective mode dispersion for composite
  fermions at $\nu=1/3$ was calculated  numerically using
  the methods in ~\cite{fett}.  This work predicted an instability
  region in qualitative agreement with the one obtained here.
  In place of a magneto-roton minimum. the squared frequency crossed the
  zero frequency axis as  the momentum grew from zero,
   and later became  positive  at  higher momenta.

  In ~\cite{w2}, we also considered the collective
   mode dispersion by using a phenomenological ansatz for the
   exact one electron Green function. It was also assumed  that because
   the mean spatial  separation between electrons is of the order of the
   expected size of the collective mode, screening effects
   in the interaction kernel of the Bethe-Salpeter equation were small.
   A   region similar to the one found in ~\cite{w1}
   and in the present work was obtained.

  It should be emphasized  that the presence of an instability for
  these fractions  does not necessarily mean that the
  Laughlin or composite fermion descriptions of
 these  systems should be limited.
  These instabilities could be necessary  for
  completing the picture given by the composite fermion description
   in which  complementary flux quanta are associated
   with fermions in spatially wide approximate one particle
   Wannier states. This point of view was
   advanced in previous works ~\cite{c1},~\cite{c2},~\cite{c3}
   by one of the authors (A.C).
   However, its validity needs to be supported by the evaluation
    of ground state energies competitive with the ones corresponding
    to the Laughlin or composite fermion states.

Moreover, other developments in the literature seem to support
the spontaneous breaking of  continous translation invariance
in FQHE ground states. One which
 argues for a lower energy in periodically perturbed variants
of the Laughlin wavefunction at $\nu=1/3$ is discussed in ~\cite {hakim}.
The presence of higher that expected peaks in numerical
evaluations of the static structure factor was also  treated in
~\cite{chui}. Another interesting result argued for
 the compatibility of the  quantized Hall effect
with the presence of crystalline structure in
 Quantum Hall crystals ~\cite{axel}.
 More recently ~\cite{stamp}, the possibility of an
  instability in the
 composite fermion description has been suggested based
 on  calculations of the
  effective mass for these particles.

Special consideration requires the  numerical evaluations for
small systems~\cite{piet}. It is clear
  that such calculations are essentially
exact. It is also true that these evaluations  predict
a ground state separated from the first excited level by a gap
in qualitatively agreement with the value
 predicted by the SMA.
Therefore, we conclude that numerical results for small systems
 do not  support any strong spontaneous
  breaking of  translation symmetry in the ground
state. However, numerical calculations do not  seem
 to be able  to rule out
the presence of a weak charge density or  correlation periodic structure
which  could affect the gap at various filling factors
through natural Bragg reflection processes.  The highest particle number
 considered up to now is near  twelve.
 With such a small number of particles,  weak periodic properties
 present
  in the ground state in the thermodynamic limit, could be hidden by
  boundary problems. It is possible that the calculated energy gap between the
  ground state and the first excited level  could also be viewed
  as arising from a
  weakly periodic state in the thermodynamic limit.

 From the experimental side,  it is well known that
 finding traces of even the existence of the Wigner crystal phase
with its strong charge density modulation is difficult. Therefore,
the presence of a weak modulation in the properties
 of the ground state seems to be, at  present, undetectable.

In the next section, the stability criterion is
given, and in Section 3 its application to  the SMA is
considered.



\section{Stability Condition}

Let us consider the following  hamiltonian describing
  2D-electrons with coulomb interactions
\ba
\hat{H}&=& \int d^2x \phx \hat{O}\px
        +\int d^2x (e \Psi^{+}(\xv) \Psi(\xv)-\deo) \fx  \nn \\
      & &  \int d^2x d^2y \Psi^{+}(\yv) \Psi^{+}(\xv) {e^2 \over
       \epsilon |\xv-\yv|}  \Psi(\xv) \Psi(\yv).
\ea

\noindent
The $ \Psi$ fields   satisfy the usual
equal time  commutation relations
 \ba
 \left[\px ,\phy \right]_{+}=\delta^{(2)}(\xv-\yv),
 \ea
\noindent
 which become in the first Landau level approximation of the QHE
 problem
\ba
 \left[\px ,\phy\right]_{+}=\Pi_{0}(\xv-\yv),
 \ea

\noindent
for the  specific fractional Hall effect system to be
considered below . In (3), the $\Pi_{0}$ function
is the projection operator onto the first Landau level.
 In this  case, the free
particle kinetic energy term determined by $\hat{O}$ in  (1) will be
absent as usual because the first Landau level restriction makes it a
constant energy shift.  A uniform background charge density $\rho_o$
is assumed as usual to assure the stability of the system without
breaking translation invariance. A macroscopic
 dielectric constant $\epsilon$
is also considered for the medium in
 which the electronic system is imbedded.

The 2D-electron gas interacts
with an electrostatic field $\fx$ which is  created by
arbitrary but small fluctuations in the charge
 density of the background
$\rho_f(\xv)$. The field $\fx$ is therefore linked to $\rho_f(\xv)$
through
\ba
\fx=\int { 4 \pi \over \epsilon |\xv-\yv|}\, \rho_{f}(\yv)\, d^3 y.
\ea

Assuming that the 2DEG is  at a finite temperature determined by
the parameter $ \beta=(kT)^{-1},$ the total internal energy of the
electronic system plus the fluctuating electrostatic field is given
by
\ba
U= <\hat{H}> + \int d^3x\, {\epsilon \over 8 \pi} \,  \partial \phi(\xv)
         \partial \phi(\xv),
\ea

\noindent
where the statistical mean is   considered
 in terms of the Bloch density matrix $\hat{\rho}$  and the
partition function $Z$
\ba
<\hat{O}>= {Tr(\hat{O} \hat{\rho}) \over Tr(\hat{\rho})}, \\
\hat{\rho}= \exp (-{\hat{H} \over kT}), \\
Z=Tr(\hat{\rho}).
\ea

After using the  expression for the entropy ~\cite{lanstat}
\ba
S=-<ln {\hat{\rho} \over Z} >={1 \over kT}<\hat{H}>+kT ln Z,
\ea

 \noindent
 the total  Free-Energy of the electronic gas plus electrostatic
 fluctuations  can be written in the following form
\ba
\Gamma[T,\phi]=U -T S= \Gamma_{e}[T,\phi]+ \int dx^3 \, {\epsilon \over 8 \pi}
\,  \partial \phi(\xv) \partial \phi(\xv),
\ea

\noindent
where the electronic contribution $\Gamma_e$ is given by
\ba
\Gamma_{e}[T,\phi]=-{1 \over \beta} ln[ Tr (\exp(-\hat{H} \beta))].
\ea

The starting point of our analysis  will be the
condition that the work
 needed to be done at constant temperature  to produce
the arbitrary  background charge density fluctuation $\rho_f(\xv)$
should always be positive  for the system to be stable.
Therefore, the second derivative  of the Free-Energy
 functional over the field $\phi$ should be a positive
definite kernel.  Taking the second derivative of (10) gives the
following relation
\ba
\ga |_{\phi=0}&=&-{\epsilon \over 4 \pi} \partial^2(\xv)
           \delta^{3}(\xv-\yv)+\gae |_{\phi=0}  \delta(x_3,y_3)
\ea
where
\ba
            \delta(x_3,y_3)&=&\delta(x_3) \delta(y_3).
\ea

The product of two $\delta$ functions in the coordinates
 orthogonal to the 2DEG plane appears because the functional
  derivatives over $\phi$ are taken in  3D-space while the
   electron gas is confined to the $x_3=0$ plane.

 To evaluate the electronic contribution in (12), it is convenient
to use the Matsubara approach by introducing temperature Heisenberg
field variables
\ba
\Psi(\xv,x_4)&=&\exp(x_4 \hat{H}) \Psi(\xv) \exp(-x_4 \hat{H}), \\
\Psi^{+}(\xv,x_4)&=&\exp(x_4 \hat{H}) \Psi^{+}(\xv) \exp(-x_4 \hat{H}).
\ea

In terms of the particle density operator
\ba
\rho(\xv,x_4)=\ppx,
\ea

\noindent
the second derivative of the electronic contribution to the
 Free-Energy at vanishing values of the field
$\phi$ (where the mean value of the density is homogeneous)
can be expressed in the following form
\ba
\gae&=&-{1 \over \beta}\ix \iy <{\large T}\left( e \rho(\xv,x_4) 
. e \rho(\yv,y_4) -\deo^2 \right)
>   \delta(x_3,y_3)       \nn \\
    &=& -\ix < e \rho(\xv,x_4) . e \rho(\yv,0) -\deo^2 > 
 \delta(x_3,y_3)    \nn \\
    &=& -\ix < e \exp(x_4 \hat{H}) \rho(\xv) . 
e \exp(-x_4 \hat{H}) \rho(\xv)>
    \delta(x_3,y_3) \nn \\ & &+\deo^2 \beta  \delta(x_3,y_3).
\ea

The energy eigenfunctions
of the  hamiltonian (1) satisfy
\ba
\hat{H} |n>= \enen |n>
\ea

\noindent
and a completeness relation for them can be introduced
after the density operator
$\rho(\xv)$ in (17). This expression becomes
\ba
\gae &=&-{e^2 \delta(x_3,y_3) \over Z} \sum_{n,m\neq 0}
 {\exp[(-\enem \beta)-\exp(-\enen \beta) \over \enen-\enem }
  <n|\rho(\xv)|m><m|\rho(\yv)|n> \nn \\
   & &- {e^2 \delta(x_3,y_3) \over Z}
   \beta <0|\rho(\xv)|0><0|\rho(\yv)|0>
   +\deo^2 \beta \delta(x_3,y_3).
\ea

Let us consider now the zero temperature limit.
 As the temperature approaches  zero the second term
in  (19) tends to cancel the constant
contribution which is proportional
to $\rho_0^2$ since the partition function $Z \rightarrow 1$
(a non degenerate ground state is assumed) and the
mean value densities are constant being
 balanced by the background jellium.
In the remaining sum, only terms having one of the indices
 $ n$ or $m$  zero will survive
 if the ground state energy is nondegenerate.
 Therefore, (19) reduces to
\ba
\gae {\large |}_{T \rightarrow 0}=-e^2 \delta(x_3,y_3) \sum_{n \neq 0}
                       {2 Re[<0|\rho(\xv)|n><n|\rho(\yv)|0>]
                        \over \enen -\epsilon_{0} }.
\ea

After substituting (20) in (12), the second derivative of the
total Free-Energy  in the zero temperature limit takes the form
\ba
\ga |_{T \rightarrow 0} &=&-{\epsilon \over 4 \pi} \nabla^2_{\xv}
       \delta^{(3)}(\xv-\yv)-  \nn \\
   & &    \delta(x_3,y_3) \sum_{n \neq 0} e^2
       {2 Re[<0|\rho(\xv)|n><n|\rho(\yv)|0>]
                     \over \enen -\epsilon_0}.
\ea

We restrict in what follows  the fluctuation
 charge densities  to those  that are
non vanishing only in the 2DEG plane $x_3=0$. In this case the
associated electrostatic field $\phi$ can be expressed
in the Fourier representation
\ba
\fx=\int{d^2k \over (2 \pi)^2}\,  \fk \, \exp(i\kv.\xv-|\kv||x_3|),
\ea

\noindent
where $\fk$ is any complex function of the momentum variables
$\vec{k}$ in the plane satisfying $ \fk^*=\phi(-\kv)$.

The above property follows from acting with the 3D-laplacian
on (22) which produces Poisson equation
\ba
\nabla^2\fx=-\int{d^2k \over (2 \pi)^2} 2 |\kv| \, \fk \, \exp(i\kv.\xv)
\,\, \delta(x_3)=-4 \pi \rho_{f}(\xv),
\ea

\noindent
where $\rho_f(\xv)$ is the assumed  arbitrary surface
charge density fluctuation.

Using  this class of potentials allows us to write the
quadratic form
\ba
\Gamma^{(2)}[\phi]&=& \int d^3x d^3y \, \fx \ga \fy    \nn \\
          &=&\int {d^2 k \over (2 \pi)^2 } \left( {2 |\kv| \epsilon
             \over 4 \pi}-{ \it K} (\kv) \right) \phi^{*}(\kv)\fk,
\ea

\noindent
where the function ${\it K}$ is given by
\ba
{\it K}(\kv)={2 \over V} \sum_{n \neq 0} {|<0|\rho_{\kv}|n>|^2
             \over \enen-\epsilon_o}
\ea
and $\rho_{\vec{k}}$ is the 2D-Fourier transform of the density operator.

Since  the quadratic form (24) must be positive
 the  following condition for the stability of the
electronic gas  arises
\ba
1-{2 \pi e^2 \over \epsilon \hbar |\kv|} {2 \over V} \sum_{n \neg 0}
        {|<0|\rho_{\kv}|n>|^2 \over \enen-\epsilon_o} \geq 0.
\ea

This relation can be expressed in terms of  the electron density
$\rho=N/V$, the dynamical structure factor
\ba
S(\kv,w)={1 \over N} \sum_{n \neq 0} |<0|\rho_{\kv}|n>|^2
                  \delta(w- (\enen-\epsilon_o)/\hbar),  \,\,\, (w>0)
\ea

\noindent
and the interaction potential
$$v(\kv)= {2 \pi e^2 \over \epsilon \kv},$$
resulting in the final expression
\ba
1- 2 \rho\, v(\kv) \int^{\infty}_{0}{dw \over w} S(\kv,w) \geq 0.
\ea

\section{Analysis of the SMA }

In this section,  the  condition (28) will used to considered
 the Girvin, MacDonald and Platzman (GMP) single mode
approximation (SMA) for  collective modes in
the FQHE ground states~\cite{gmp}.
For this purpose, it is convenient to introduce the  dimensionless
wavevector
$$\qv=r_o \kv$$
and measure the energy in units of
${e^2 \over \epsilon r_o}$
where the magnetic radius is given in terms of the
magnetic field intensity  by
$r_o=\sqrt{\hbar c \over |eB|}.$

  As  discussed in ~\cite{gmp}, the SMA is characterized by
  a dynamic structure factor of the form
\ba
S(\qv,w)=\overline{s}(\vec{q}) \, \delta(w-\Delta (\vec{q})),
\ea
\noindent
where $\overline{s}$ is the static structure factor
and $\Delta(\vec{q})$ is the collective mode energy as a
function of the magneto-roton  momentum.

 In terms of the polarizability parameter
\ba
\alpha(\qv)={\overline{s}(\qv) \over \Delta (\qv)},
\ea
\noindent
and  the electron  density
filling factor $\nu$:
$$
\rho={\nu \over 2 \pi r_{o}^2},
$$
the stability condition for the approximate FQHE
states described by the SMA can be written as
\ba
{|\qv| \over 2} - \nu \alpha(\qv) \geq 0.
\ea

In order to investigate  the validity of (31) for the
SMA as applied to the main series of the $\nu=1/m$
Laughlin states,  we used the data for the polarizability
parameter obtained in ~\cite{gmp} for m=3,5,7,9.
The results indicate that
the bound (31) is violated in a finite neighborhood of
 the maxima for the polarizability parameter as a
  function of the wavevector for all these states.
The instability regions for  $\nu=1/3$ and
$\nu=1/5$  are shown in Figs. 1 and 2 as the  zones
of the curves lying above the straight lines.
These lines are
 plots of the first term in (31) after multiplying
by the factor m in each case.
Similar results are valid for the $\nu=1/7$ and $\nu=1/9$ states.
It can be seen that the instability is stronger
 for lower filling factors, as should be expected.

In conclusion, a criterion that the static dielectric
response must satisfy for an interacting
 many body theory to be translational
invariant was presented. From it, indications of an instability
in the static response of the Laughlin states  within
the SMA approximation were obtained. These results suggest the
existence of states different than the SMA variational states
which could break
continuous translation invariance and  have  energies lower than
 the Laughlin wavefunctions. Such wavefunctions could be similar to
those discussed in ~\cite{hakim} which we  suspect  are related to the
  special Hartree-Fock
 states discussed in  ~\cite{c1},~\cite{c2} and ~\cite{c3}.
 Alternatively, improvements
 of the original variational calculation of Girvin, MacDonald
 and Platzman may be possible.
 These problems will be
 considered in future work.

  The relation obtained here may also be
  useful for other problems
  where the stability of an homogeneous
  many body system under  spontaneous breaking of  translation
  symmetry may be an issue.

\section{Acknowledgments.}

One of the authors (A.C.) greatly acknowledge the support of the
High Energy Section of the ICTP for a helpful visit to the Centre
and of the Condensed Matter Section for the nice opportunity to 
participate in the Adriatico Research Conference : "The Electron 
Quantum Liquid in Systems of Reduced Dimensions"(2-5 July 1996).
The helpful remarks and comments of J. Strathdee, E. Tossati, S. Fantoni,
G. Mussardo, J. Lawson, N. H. March, A. H. MacDonald, F. Ortolani,
G. Fano and  G. Baskaran are also greatly acknowledged.

We all are also deeply indebted  to the Third World Academy of
Sciences for its support through  the TWAS Research Grant
93 120 RG PHYS LA.

\newpage

{\large \bf Figure Captions}

\begin{itemize}
\item{\bf Fig. 1}\,  A plot of the SMA polarizability parameter as a
 function of the wavevector for $\nu=1/3$ Laughlin state.
The  momentum region in which the stability condition is not
satisfied  corresponds to the zone in which
the curve is over the straight line. \\

\item{\bf Fig. 2}\, The  plot of the SMA polarizability parameter
 for the $\nu=1/5$ Laughlin state. The instability region is
 indicated by the wavevector neighborhood for which the curve
 passes  over the straight line.  \\
\end{itemize}

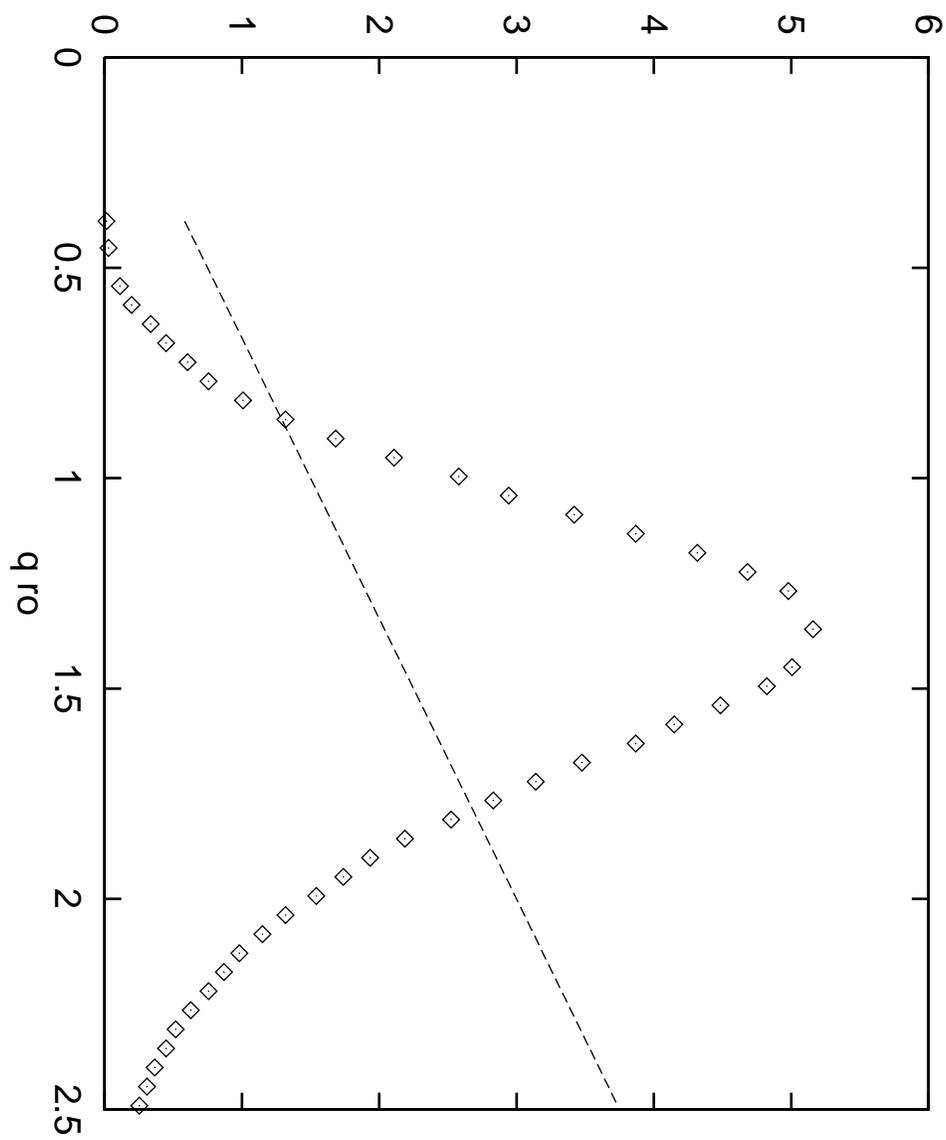
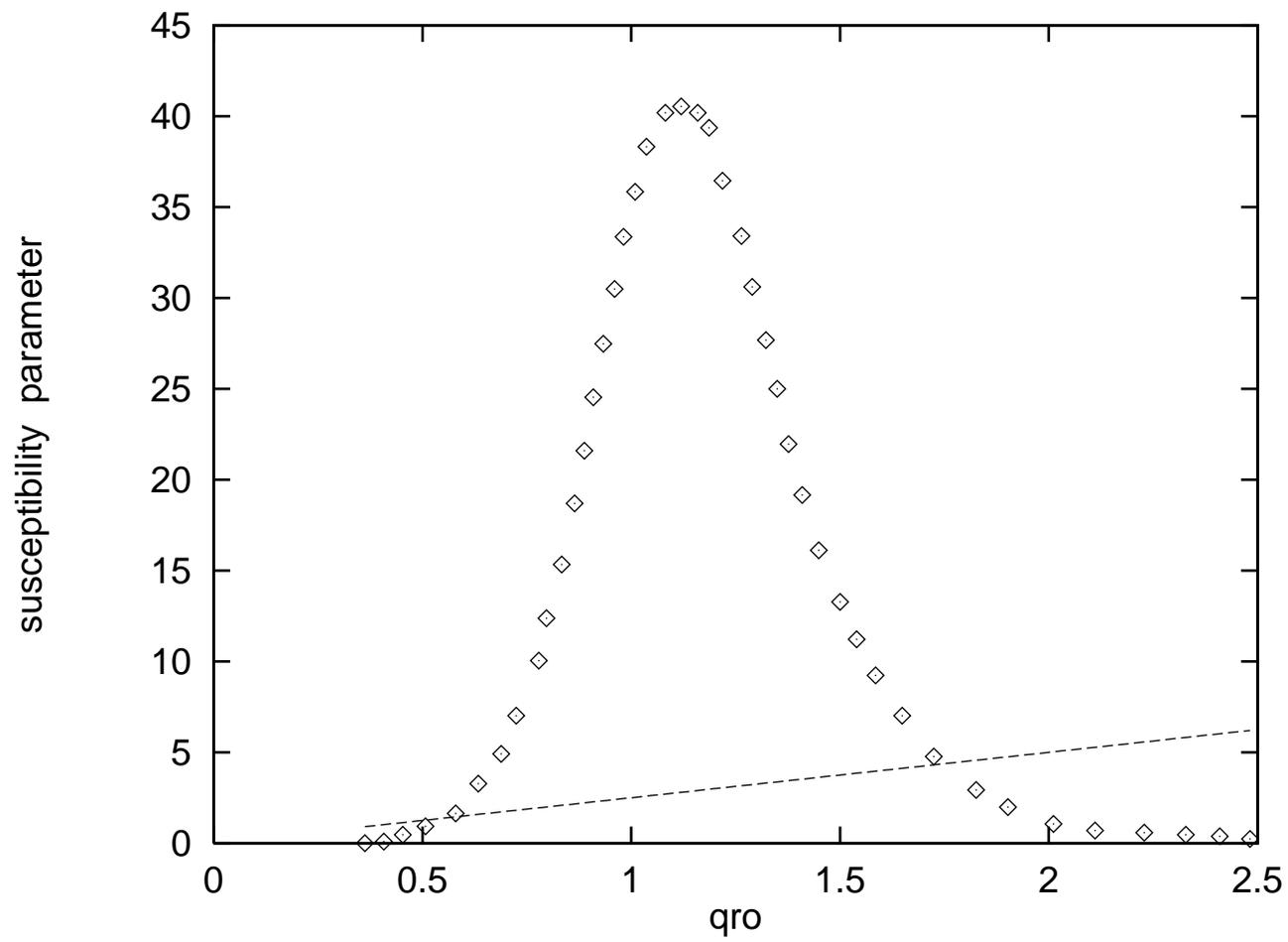
\end{document}